\documentclass[preprint,12pt]{elsarticle}

\usepackage{amssymb}

\journal{Applied Surface Science}

\begin{document}

\begin{frontmatter}



\title{A perspective on effective medium models of thermal conductivity in 
(ultra)nanocrystalline diamond films}


\author{Titus Sandu}

\address{National Institute for Research and Development in Microtechnologies-IMT, 126A, Erou Iancu Nicolae Street, Bucharest, Romania}

\author{Catalin Tibeica}

\address{National Institute for Research and Development in Microtechnologies-IMT, 126A, Erou Iancu Nicolae Street, Bucharest, Romania}
\begin{abstract}
Thermal conductivity of nanocrystalline and ultra-nanocrystalline films is 
analyzed with effective medium theory (EMT) models. The existing EMT models 
use the spherical inclusion approximation. Although this approximation works 
quite well it is inconsistent, mostly with respect to the maximal packing of 
74{\%}, which may be unrealistic for polycrystalline films. To check the 
consistency of these models we devise an EMT model with arbitrarily shaped 
inclusions. We pick the EMT model with cubic inclusions and we compare its 
results with the results of the EMT model with spherical inclusions. It is 
found a very good agreement between both calculations. This agreement is 
explained by general geometrical arguments. We further employ these models 
to analyze thermal conductivity of nanocrystalline and ultra-nanocrystalline 
diamond films. It is noticed that the effective conductivity is strongly 
affected not only by the boundary Kapitza resistance but also by intra-grain 
scattering for grain sizes below 100 nm. 
Generally, both intra-grain conductivity and Kapitza resistance 
increase with grain size. However, the effect of Kapitza resistance increase 
is negligible due to the geometrical factor accompanying Kapitza resistance 
contribution to the effective conductivity. 
\end{abstract}




\end{frontmatter}


\section{Introduction}
\label{A}
Thermal properties of nanocrystalline materials are of great interest in 
fields like micro/nanoelectromechanical systems (M/NEMS) or thermal 
management of electronics for thermal interface materials \cite{1}. In 
particular, nanocrystalline materials like nanocrystalline diamond (NCD) and 
ultrananocrystalline diamond (UNCD) films are also ideal candidates for 
mechanical and electronic applications with outstanding properties \cite{2}. 

The NCD films are defined as polycrystalline materials containing grains with 
sizes between tens of nm and 1 $\mu $m, while the UNCD films contain grains 
having sizes between 2 to 10 nm \cite{3}. On the other hand, the allotrope forms 
of carbon such as diamond, graphene, or nanotubes are the best ever known 
heat conductors, with their heat conductivity an order of magnitude larger 
than any metal \cite{2}. In a bulk material thermal conductivity is given by the 
following equation obtained from a kinetic theory: \textit{$\kappa $}=$ C_{V}V_{g}$\textit{$\Lambda $}/3. 
$C_{V}$ is the heat capacity, $V_{g}$ is the average phonon group velocity, 
and \textit{$\Lambda $} is the phonon mean free path. However, thermal conductivity of 
nanocrystalline materials and of NCD and UNCD films, in particular, are 
considerably reduced due to grain boundary effects which add a boundary 
thermal resistance and reduced mean free paths of phonons from their bulk 
values \cite{4,5}. 

\textit{Generally, thermal conductivity of nanocrystalline materials decreases with 
the decrease of crystallite size} \cite{4, 5}. Microscopic techniques such as Boltzmann transport 
equation \cite{6} and Monte Carlo simulations \cite{7} can consider detailed phonon 
grain boundary scattering but they are numerically intensive and also expensive. 
Another microscopic approach is the phonon-hopping model \cite{8}. The model 
assumes that in a nanograin the phonon transport is governed by the same 
mechanisms as in bulk crystal and introduces a phonon hopping parameter, 
which describes the phonon transport from one grain to the next one. Although 
the phonon-hopping model is able to predict the temperature dependence of 
the effective thermal conductivity \cite{9}, one can be easily show that the 
model is equivalent to the effective medium theory (EMT) of Yang \textit{et al.} \cite{10}. 
Furthermore, even a combined serial/parallel heat conduction model like that used to 
explain thermal conductivity in NCD films \cite{23} can be cast into the form given in \cite{10}.

The EMT model of Yang \textit{et al.} is based on the EMT of Nan and Birringer, who 
considered spheres as shapes for crystallites \cite{4}. However, the maximal 
packing obtained with equally-sized spheres is 74{\%}, therefore one may 
encounter some inconsistencies of the model based on spherical crystallites. 
In addition, spherical crystallites are inconsistent with the 
assumption of planar interfaces between grains made in the model described in 
ref.~\cite{10}. Moreover, in the EMT model of Yang \textit{et al.} the grain conductivity is 
simply the bulk conductivity. This model can be further improved by adjusting the phonon mean free path 
of the grain to the grain size \cite{8,9}. Recently, Dong \textit{et al.} refined the model of Yang \textit{et al.} by 
considering an intra-grain conductivity that is different from the bulk 
value and is affected by the phonon scattering at the grain boundary \cite{11}. 

In this work we analyze the effective thermal conductivity of 
nanocrystalline and ultra-nanocrystalline films with various EMT models. We relate our 
recent results regarding the homogenization of heterogeneous materials \cite{13} 
to the EMT models expressed in Refs. \cite{4,8,9}. We devise an EMT 
with arbitrarily shaped crystallites and we compare an EMT model with 
cubic-shaped crystallites in 3D and square-shaped crystallites in 2D with 
the model of Yang \textit{et al.}~\cite{10}. We further analyze the 
influence of various factors like crystallite size, phonon scattering at the grain boundary, 
and boundary thermal resistance on the thermal conductivity of NCD and UNCD 
materials. Our paper is structured as follows. In the next section we 
present the effective medium models of thermal conductivity in 
polycrystalline materials and our results regarding the EMT model with arbitrarily 
shaped crystallites. In section 3 we present numerical 
results and our analysis regarding thermal conductivity of NCD and UNCD 
materials. The last section is dedicated to conclusions.

\section{Effective Medium Models of thermal conductivity in nanocomposites and polycrystalline materials}
\label{B}

In the modeling of composites the electric, dielectric, or thermal phenomena 
can be treated similarly, since the governing equations are the same \cite{14}. 
Moreover, if the composite has a granular structure (a collection of 
inclusions dispersed into a matrix), the calculations of the composite 
effective parameters can start from the behavior of a single inclusion 
embedded in the matrix. Having the response of a single inclusion and 
considering the interactions between inclusions, the effective parameters of 
the composite can be calculated with various degrees of accuracy \cite{14}. This 
process may be called a homogenization process by which the response of a 
heterogeneous system is similar to that of a homogeneous system characterized 
by the effective parameters calculated in the process. The 
response of a particle of volume $V_{1}$ immersed in a uniform matrix and 
exposed to a uniform heat flux can be defined as \cite{13}

\begin{equation}
\label{eq1}
\alpha = \frac{1}{4\pi V_1 }\int\limits_{V_1 } {\frac{\left( {\kappa _1 - 
\kappa _0 } \right)}{\kappa _0 }\left( { - \nabla T\left( r \right)} 
\right)d^3r} .
\end{equation}

In Eq. (\ref{eq1}) $\kappa _1 $ and $\kappa _0 $ are the conductivities of the 
particle and of the embedding medium, respectively and $T$ is the temperature 
field calculated in the presence of the particle. In dielectric 
phenomena \textit{$\alpha $} is the normalization polarizability which can be cast as in the 
following \cite{13,15}

\begin{equation}
\label{eq2}
\alpha = \sum\limits_k {\frac{w_k \left( {\kappa _1 - \kappa _0 } 
\right)}{\kappa _0 + L_k \left( {\kappa _1 - \kappa _0 } \right)}} ,
\end{equation}

\noindent
where $L_{k}$ is the depolarization factor of the $k^{th}$ eigenmode of an 
electrostatic operator defined on the boundary of $V_{1}$ and $w_k $ is the weight of this eigenmode. It can be readily seen 
that \textit{$\alpha $} is related to the $t$-matrix formalism of the scattering theory \cite{14,16}. 
In addition, the interfacial barrier resistance between the inclusion and 
the matrix is modeled as a very thin shell of thermal conductivity $\kappa 
_2 $ \cite{4,10,14}. The shelled particle can be also ``homogenized'' by simply 
replacing it with a similar particle, which has an effective conductivity 
for each eigenmode given by \cite{13,15}:

\begin{equation}
\label{eq3}
\kappa _{k\_1} = \kappa _2 \left( {1 + \frac{\kappa _1 - \kappa _2 }{\left( 
{1 + \eta } \right)\kappa _2 + \eta (1 / 2 - \chi _k )\left( {\kappa _1 - 
\kappa _2 } \right)}} \right),
\end{equation}

\noindent
where $1 + \eta $ is a volume ratio between the particle with and without the 
shell. Equation (\ref{eq3}) is a generalization of the results from \cite{4,17} to 
arbitrarily shaped nanoparticles. It is easy to check that interfacial 
resistance is obtained in the limit $\eta \to 0$ and $\eta \mathord{\left/ 
{\vphantom {\eta {\kappa _2 }}} \right. \kern-\nulldelimiterspace} {\kappa 
_2 } \to $ constant as

\begin{equation}
\label{eq4}
\kappa _{k\_1} = \frac{\kappa _1 }{1 + {d\kappa _1 L_k R_K } \mathord{\left/ 
{\vphantom {{d\kappa _1 L_k R_K } l}} \right. \kern-\nulldelimiterspace} 
l},
\end{equation}

\noindent
where $d$ is the dimension of the problem, which is 3 for a 3D problem and 2 
for a 2D problem, $R_{K}$ the is the Kapitza boundary resistance and $l$ is a 
characteristic length of the particle. In Eq. (\ref{eq4}) we considered that the 
grain boundary is shared between two adjacent grains as in the paper of 
Yang\textit{ et al.} \cite{10}.

Once the response of a single inclusion is calculated various procedures 
can be devised in order to obtain effective medium models. One of the oldest 
and the most successful effective medium models is the Maxwell-Garnett 
approximation, which is an averaged $t$-matrix approximation \cite{14,16}. The 
Maxwell-Garnett approximation works well for low volume concentrations of 
the filler \cite{13}. For large volume concentrations, however, the coherent 
potential approximation (CPA) is more appropriate \cite{13,14}. 

\begin{figure}
\centerline{\includegraphics[width=3.0in]{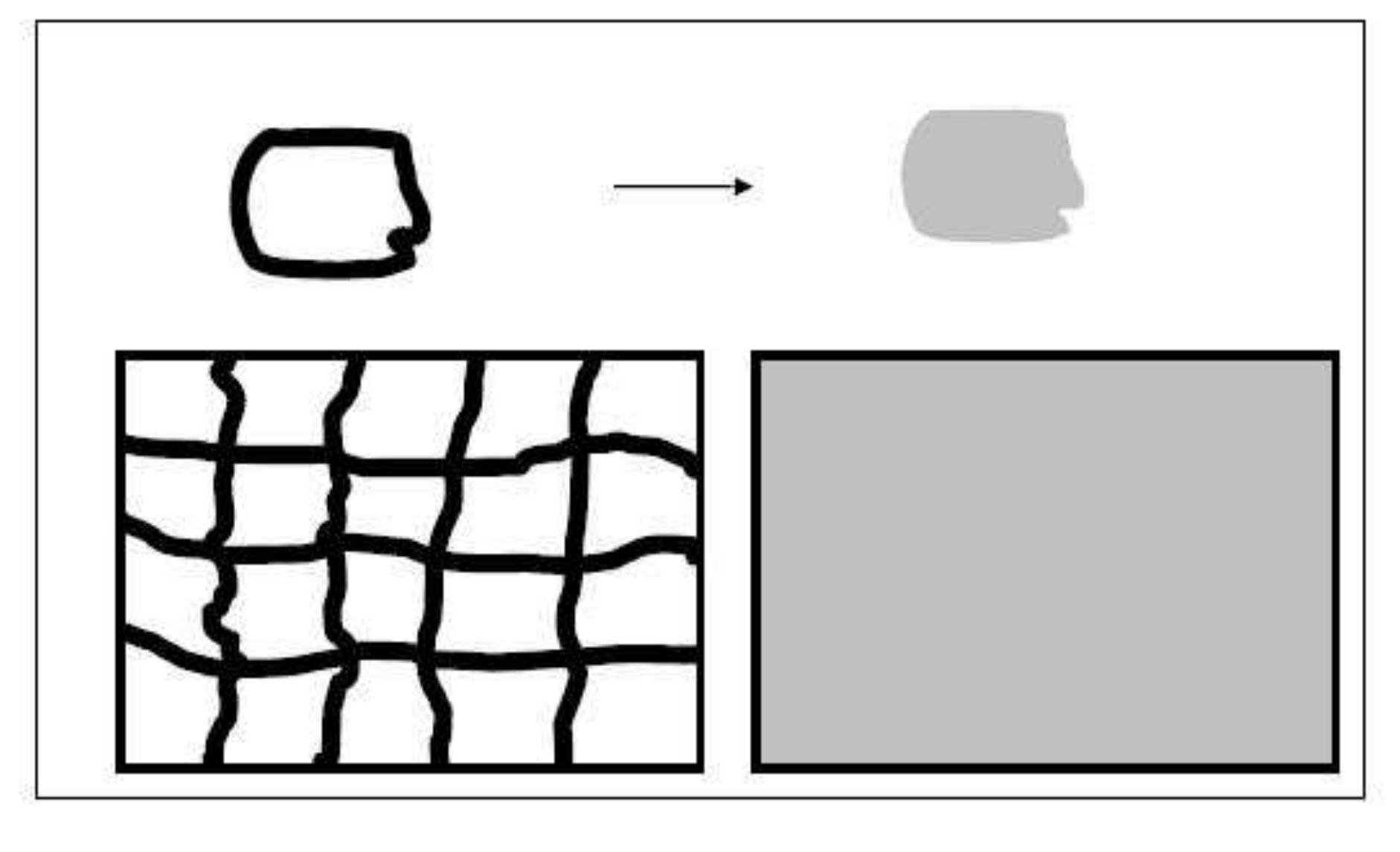}}
\caption{Process of homogenization in nanocrystalline materials.  It
  is represented by the homogenization of a grain and its boundary
  (top), followed by the homogenization of the whole material. The
  thick black lines represent the grain boundaries.}
\label{fig1}
\end{figure}

In CPA, the effective conductivity $\kappa _{eff} $ is obtained by setting 
the average of \textit{$\alpha $} over the composite to zero, i. e., $\left\langle \alpha 
\right\rangle = 0$. For arbitrarily-shaped inclusions the explicit form of 
CPA is \cite{13}

\begin{equation}
\label{eq5}
\sum\limits_k {f\frac{w_k \left( {\kappa _{k\_1} - \kappa _{eff} } 
\right)}{\kappa _{eff} + L_k \left( {\kappa _{k\_1} - \kappa _{eff} } 
\right)}} + \left( {1 - f} \right)\frac{w_k \left( {\kappa _0 - \kappa 
_{eff} } \right)}{\kappa _{eff} + L_k \left( {\kappa _0 - \kappa _{eff} } 
\right)} = 0,
\end{equation}

\noindent
where $f$ is the volume concentration of the inclusions. When the inclusions 
with a boundary thermal resistance fill the whole composite, i. e, $f $= 1, Eq. 
(\ref{eq5}) becomes

\begin{equation}
\label{eq6}
\sum\limits_k {\frac{w_k \left( {\kappa _{k\_1} - \kappa _{eff} } 
\right)}{\kappa _{eff} + L_k \left( {\kappa _{k\_1} - \kappa _{eff} } 
\right)}} = 0,
\end{equation}

\noindent
which can be a model for thermal conductivity of polycrystalline films
and an extension of the model used by Nan for spherical and ellipsoidal grains \cite{4}. A graphical
illustration of EMT described by Eq. (\ref{eq6}) is given in
Fig.~\ref{fig1}, where the grain with its boundary is transformed into
a homogeneous body and then the process is applied to the whole
material.

Generally, Eq. (\ref{eq6}) has as many solutions as the number of eigenmodes present 
in the response. However, just one solution has physical significance, which 
must be a positive root. A practical way to evaluate the effective thermal 
conductivity $\kappa _{eff} $ is to look for a solution that satisfies 
$\left| {\kappa _{k\_1} - \kappa _{eff} } \right| \ll \kappa _{eff} $. 
Knowing that $0 < L_k < 1$ and using the sum rule $\sum\limits_k {w_k } = 1$ 
\cite{15,18}, the following approximation can be obtained 

\begin{equation}
\label{eq7}
\kappa _{eff} = \sum\limits_k {\frac{w_k \kappa _1 }{1 + {d\kappa _1 L_k R_K 
} \mathord{\left/ {\vphantom {{d\kappa _1 L_k R_K } l}} \right. 
\kern-\nulldelimiterspace} l}} ,
\end{equation}

\noindent
where $\kappa _1 $ is the grain thermal conductivity of the grain. 
In the case of spherical grain $L_{k}$ = 1/3 and $w_{k}$ = 1 thus Eqs. (\ref{eq6}) and 
(\ref{eq7}) have the form similar to that given in Ref. \cite{10}

\begin{equation}
\label{eq8}
\kappa _{eff3D} = \frac{\kappa _1 }{1 + {R_K \kappa _1 } \mathord{\left/ 
{\vphantom {{R_K \kappa _1 } {l}}} \right. \kern-\nulldelimiterspace} 
{l}},
\end{equation}

\noindent
where $l$ is the diameter of the spherical grain. Eq. (\ref{eq8}) considers that 
the inter-grain space is shared between two adjacent grains \cite{10}. In the 
original work of Nan and Birringer (Ref. \cite{4}), the diameter $l$ was replaced by 
the sphere radius.

\section{Thermal conductivity models of NCD and UNCD films. Discussion}
\label{C}
\subsection{3D models}

The model of Nan and Birringer assumes spherical grains, but in reality the 
grains have polyhedral shapes and it is supposed that their departure from 
spherical shape is quite small. There are, at least, two major issues here 
regarding the effective medium models with spherical inclusions. The maximal 
packing of equal spheres is about 74{\%}, while for randomly arranged 
spheres the packing is even smaller \cite{12}. The second issue is about the 
argument of Yang\textit{ et al.} \cite{10}, who made the reasoning of interface thermal 
resistance for planar interfaces. Thus it is natural to ask if the shape 
averaging can be made around cubic rather than spherical shape. For cubic 
geometry the packing can reach 100{\%} and cubic shape is also compatible 
with planar interfaces encountered in polycrystalline materials. In the 
following we compare the results provided by the effective medium model 
using spherical inclusions with the effective medium model that assumes 
cubic inclusions. For cube we use the spectral parameters calculated in Ref. 
\cite{18}, namely $w_{1}$ = 0.44, $L_{1}$ = 0.214, $w_{2}$ = 0.24, $L_{2}$ = 0.297, 
$w_{3}$ = 0.04, $L_{3}$ = 0.345, $w_{4}$ = 0.05, $L_{4}$ = 0.440, $w_{5}$ = 0.01, 
$L_{5}$ = 0.563, $w_{6}$ = 0.09, $L_{6}$ = 0.706, $w_{7}$ = 0.04, $L_{7}$ = 0.33. 
In the cubical model $l$ is the edge length of the cube.

\begin{figure}[htbp]
\centerline{\includegraphics{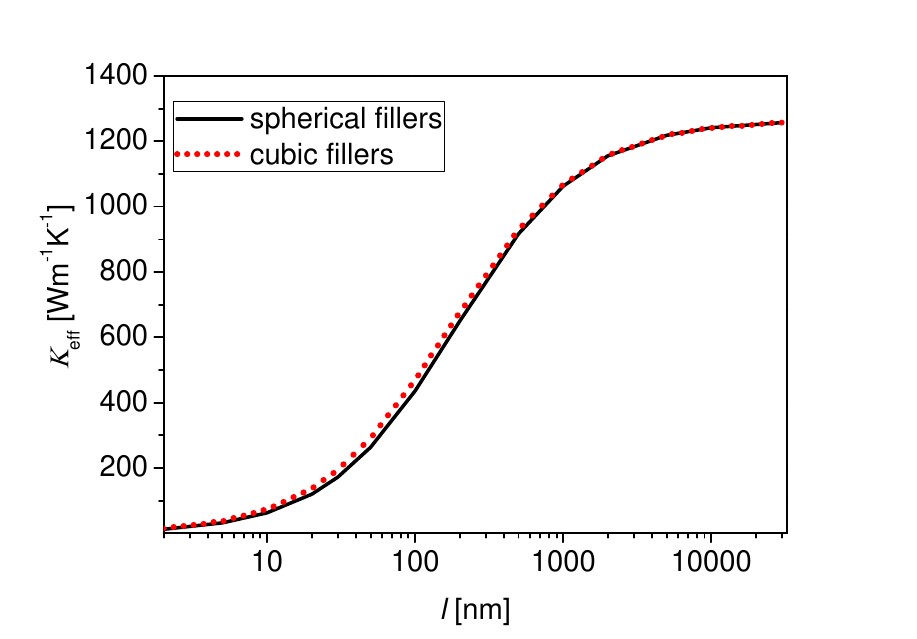}}
\caption{The effective thermal conductivities given by the EMT with 
	spherical (solid black line) versus cubic (dotted red line) shaped crystallites.}
\label{fig2}
\end{figure}



In order to assess the effectiveness of Eq.~(\ref{eq7}) we compare its outcome with the results provided by the full form of Eq.~(\ref{eq6}) for polycrystalline diamond 
using the model with cubic inclusions. The results were apart by less than 1{\%} for cubic inclusions of 
edge lengths ranging from 2 nm to 10 $\mu $m , hence Eq. (\ref{eq7}) provides reliable 
results. In Fig.~\ref{fig2} we compare the effective medium models of spherical (Eq.~(\ref{eq7})) and cubic 
inclusions (Eq.~(\ref{eq8})). The bulk thermal conductivity is 1265 Wm$^{-1}$K$^{-1}$ 
taken from molecular dynamics simulations \cite{11} and $R_{K}$=1.0 $\times$ 10$^{-9}$ 
Km$^{2}$/W. As we can see from Fig.~\ref{fig2} there is a very good match between the 
model with spherical inclusions and the model with cubic inclusions.

\subsection{2D models}

In 2D the filling factor of the plane with circles is \textit{$\pi $/}4 $ \approx $ 
78.5{\%}. Moreover, for a two-dimensional model of a polycrystalline films 
with circular inclusions, Eqs (\ref{eq4}) and (\ref{eq6}) provide a similar formula to Eq. 
(\ref{eq8}), i. e.,

\begin{equation}
\label{eq9}
\kappa _{eff2D} = \frac{\kappa _1 }{1 + {R_K \kappa _1 } \mathord{\left/ 
{\vphantom {{R_K \kappa _1 } l}} \right. \kern-\nulldelimiterspace} l },
\end{equation}

\noindent
where $l$ is the diameter of the circle. We recall that in 2D the 
depolarization factor of a circle is $L_{k}$ = 1/2 \cite{19}. 

\begin{center}
\begin{figure}[htbp]
    \centerline{\includegraphics[width=3in]{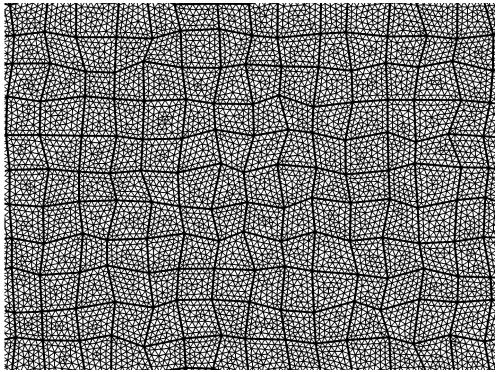}}
    \caption{A FEM simulation box to calculate the effective 
	thermal conductivity of a 2D polycrystalline material. Thick lines define 
	the grain boundaries and thinner lines show the mesh.}
    \label{fig3}
\end{figure}

\end{center}


To check the validity of Eq. (\ref{eq9}) in 2D we performed finite element (FEM) 
calculations with the commercial software COMSOL. We started with a 
simulation box that was divided in equal-sized squares which are associated 
with the crystallites. Each crystallite exhibits an interface thermal 
resistance. To mimic the random variation about square shape of the 
crystallites we moved each vertex by a random quantity between the values 
\textit{dx} and \textit{--dx} and between \textit{dy} and \textit{--dy} on $x$- and $y$-axes, respectively (for convenience we took $dx = dy$). A sample of the 
simulation box is given in Fig.~\ref{fig3}. The interface thermal resistance is 
depicted by thick separating lines. In calculations there were used the 
following values: $R_{K}$=1x10$^{ - 9}$ K$ $m$^{2}$/W for Kapitza 
interface resistance and \textit{$\kappa $}=2200 Wm$^{ - 1}$K$^{ - 1}$ for thermal conductivity.


\begin{figure}[htbp]
	\centerline{\includegraphics[width=3in,height=2in]{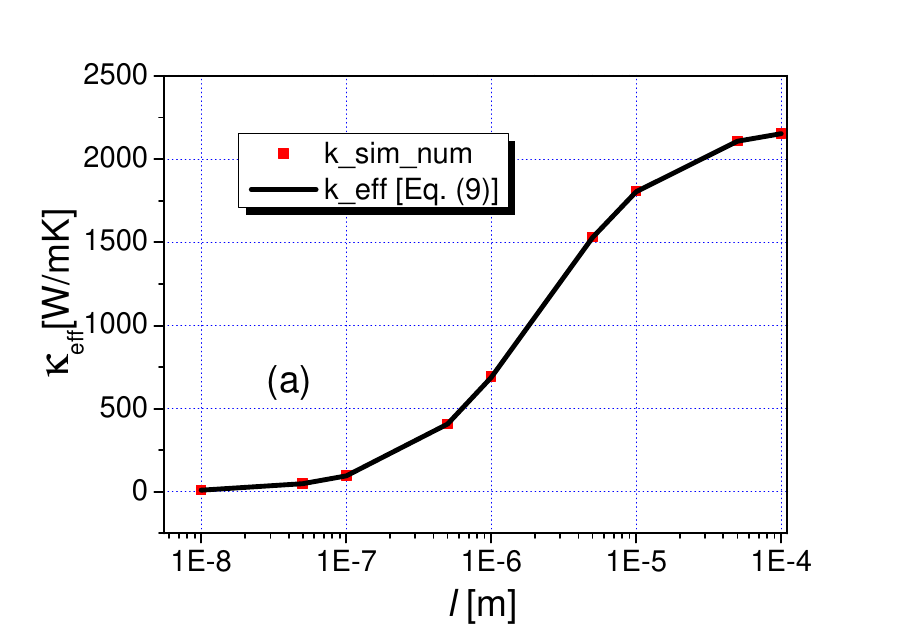}
	\includegraphics[width=3in,height=2in]{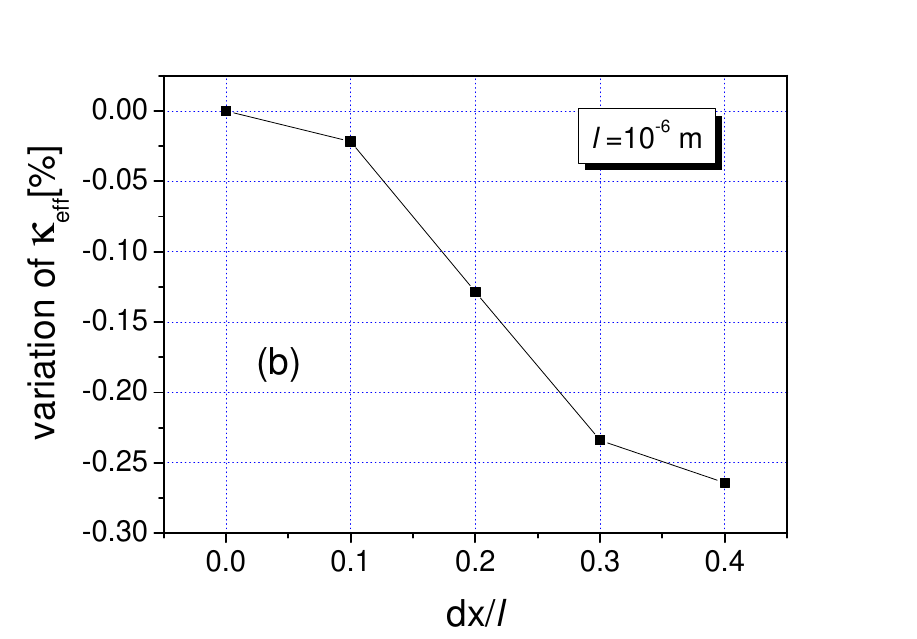}}
    \caption{(a) Comparison between calculations with a FEM method and 
    	Eq. (\ref{eq9}). The analytical results were plotted with solid black line, while the FEM results are illustrated by red square symbols; (b) The relative variation of thermal conductivity with respect to the size fluctuation \textit{dx} in FEM calculations.}
	\label{fig4}
\end{figure}


In Fig.~\ref{fig4}(a) we compare the results of FEM calculations with Eq. (\ref{eq9}) that 
gives the thermal conductivity of the EMT with circular inclusions. FEM 
calculations provide an averaging about square-shaped inclusions. Inspecting 
Fig.~\ref{fig4}(a) we may conclude that results obtained by averaging about 
circular-shaped inclusions are similar to the results obtained by averaging 
around square-shaped inclusions. In Fig.~\ref{fig4}(b) it is shown that variation of 
the results with respect to the size of \textit{dx} is far less than 1{\%}.

\subsection{The effect of intra-grain conductivity on the effective conductivity of NCD and UNCD films}

In Eqs. (\ref{eq7})-(\ref{eq9}) \textit{$\kappa $}$_{1}$ is basically the bulk thermal conductivity, which 
cannot be the intra-grain conductivity at least for grains whose sizes are 
smaller that the bulk phonon mean free path. In many calculations the phonon 
mean free path is replaced by the actual size of the grain. However, 
recently Dong \textit{et al. }found an intra-grain conductivity that includes basically all 
the intra-grain scattering events \cite{11}. By molecular dynamics simulations 
the authors obtained the intra-grain conductivity such that the final EMT 
formula for thermal conductivity in nanocrystalline materials is

\begin{equation}
\label{eq10}
\kappa _{eff} = \frac{{\kappa _1 } \mathord{\left/ {\vphantom {{\kappa _1 } 
{\left( {1 + \Lambda \mathord{\left/ {\vphantom {\Lambda {l^{0.75}}}} 
\right. \kern-\nulldelimiterspace} {l^{0.75}}} \right)}}} \right. 
\kern-\nulldelimiterspace} {\left( {1 + \Lambda \mathord{\left/ {\vphantom 
{\Lambda {l^{0.75}}}} \right. \kern-\nulldelimiterspace} {l^{0.75}}} 
\right)}}{1 + {R_K \left[ {{\kappa _1 } \mathord{\left/ {\vphantom {{\kappa 
_1 } {\left( {1 + \Lambda \mathord{\left/ {\vphantom {\Lambda {l^{0.75}}}} 
\right. \kern-\nulldelimiterspace} {l^{0.75}}} \right)}}} \right. 
\kern-\nulldelimiterspace} {\left( {1 + \Lambda \mathord{\left/ {\vphantom 
{\Lambda {l^{0.75}}}} \right. \kern-\nulldelimiterspace} {l^{0.75}}} 
\right)}} \right]} \mathord{\left/ {\vphantom {{R_K \left[ {{\kappa _1 } 
\mathord{\left/ {\vphantom {{\kappa _1 } {\left( {1 + \Lambda 
\mathord{\left/ {\vphantom {\Lambda {l^{0.75}}}} \right. 
\kern-\nulldelimiterspace} {d_1^{0.75}}} \right)}}} \right. 
\kern-\nulldelimiterspace} {\left( {1 + \Lambda \mathord{\left/ {\vphantom 
{\Lambda {l^{0.75}}}} \right. \kern-\nulldelimiterspace} {l^{0.75}}} 
\right)}} \right]} {l}}} \right. \kern-\nulldelimiterspace} {l}}.
\end{equation}
In Eq. (\ref{eq10}) \textit{$\kappa $}$_{eff}$ and \textit{$\kappa $}$_{1}$ are given in Wm$^{ - 1}$K$^{ - 1}$, the bulk 
phonon mean free path $\Lambda $ and the size of the grain $l$ are given in 
nm, while $R_{K}$ is provided in m$^{2}$KW $^{ - 1}$. Dong \textit{et al.} calculated a 
diamond bulk thermal conductivity of 1265 Wm$^{ - 1}$K$^{ - 1}$ and a 
diamond phonon mean free path $\Lambda $ of 180 nm. 


\begin{figure}[htbp]
		\centerline{\includegraphics[width=3in,height=2in]{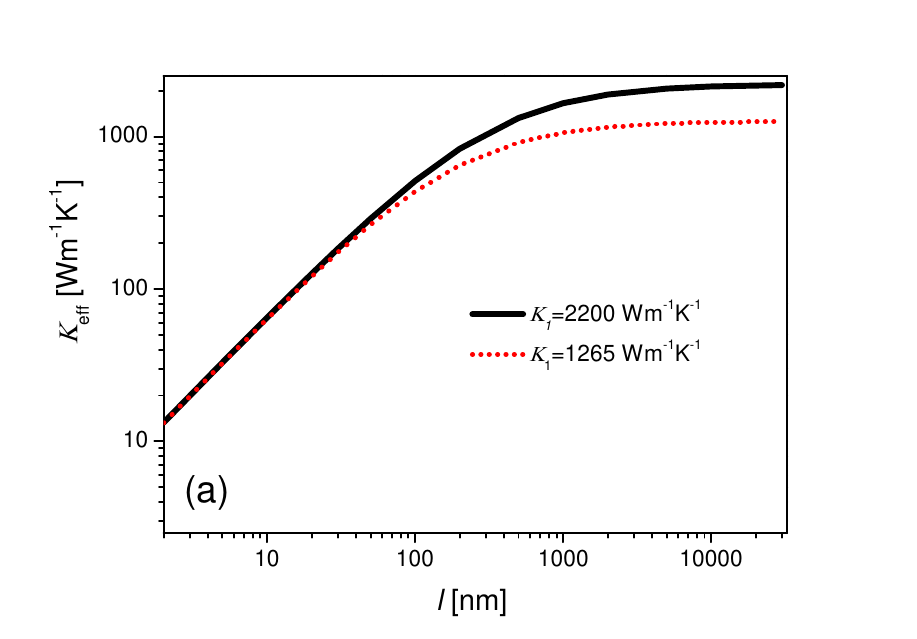} \includegraphics[width=3in,height=2in]{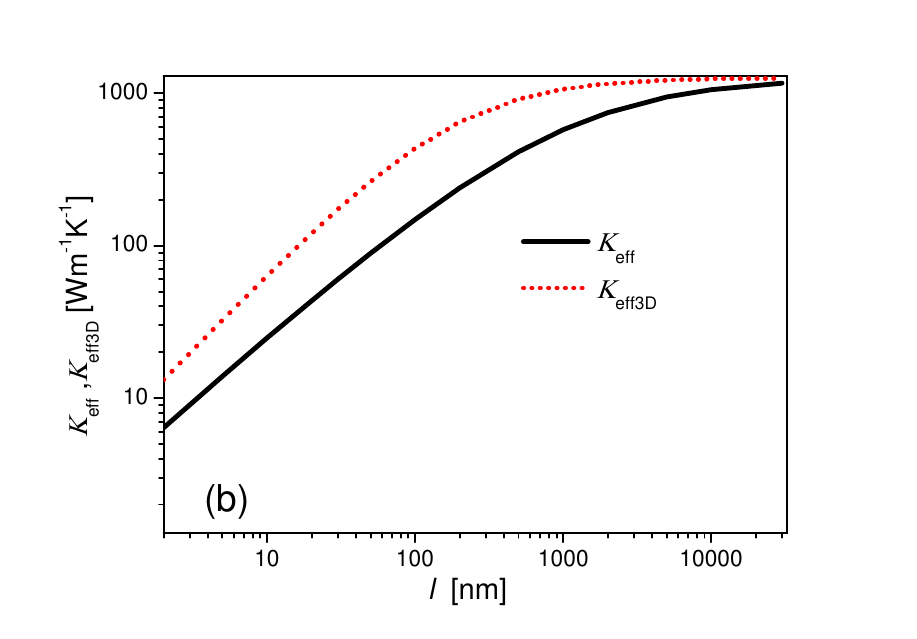}}
	    \caption{(a) The effect of the bulk thermal conductivity on the 
	    	effective conductivity of NCD and UNCD materials modeled with Eq. (\ref{eq8}); (b) 
	    	The effect of intra-grain scattering by comparing Eqs. (\ref{eq8}) and (\ref{eq10}). The 
	    	plots are given as a function of the grain size.}
	\label{fig5}
\end{figure}


In Fig.~\ref{fig5}(a) we have analyzed Eq. (\ref{eq8}) by studying the effect of bulk 
thermal conductivity on the thermal effective conductivity of UNCD 
materials. We considered two values of bulk crystalline diamond: the 
measured value of 2200 Wm$^{ - 1}$K$^{-1}$ \cite{1,2} and the calculated 
value of 1265 Wm$^{ - 1}$K$^{ - 1}$ \cite{11}. The value of Kapitza resistance 
was a typical one of $R_{K}$ = 0.15x10$^{ - 9}$ Km$^{2}$/W. We can 
easily notice that for grain sizes up to 100 nm the effective thermal 
conductivity is insensitive to the thermal conductivity of the crystalline 
bulk diamond. This is valid as long as ${R_K \kappa _1 } \mathord{\left/ 
{\vphantom {{R_K \kappa _1 } l}} \right. \kern-\nulldelimiterspace} l \gg 1$ 
according to Eq. (\ref{eq8}). 

In Fig.~\ref{fig5}(b) we compare the effect of intra-grain 
conductivity/scattering on the effective conductivity of NCD and UNCD 
materials by comparing Eq. (\ref{eq10}), where intra-grain scatterings are 
considered, with Eq. (\ref{eq8}), where such effects are neglected. We notice that 
the curves have similar shapes. Nevertheless, intra-grain scatterings lower 
the effective conductivity of 
polycrystalline diamond for the whole range of crystallite sizes from 2 nm to 
10 $\mu$m. 



	\begin{figure}[htbp]
		\centerline{\includegraphics{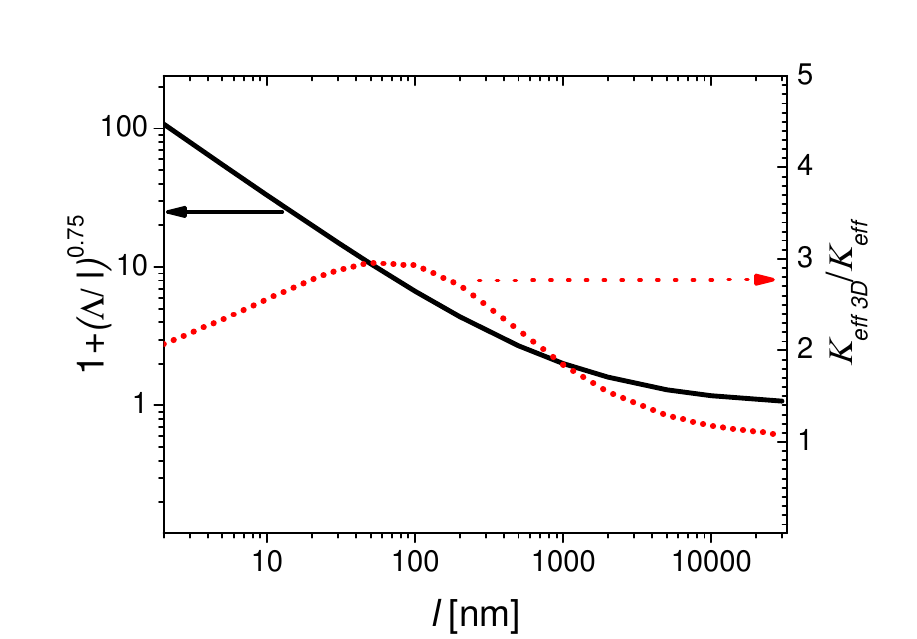}}
	    \caption{The grain-size dependence of the term (1+ \textit{$\Lambda $/l}$^{0.75})$ 
	    	responsible for the intra-grain scattering on the left-hand side $y$-axis of the graph (solid black line). On the right-hand side $y$-axis of the graph is illustrated the ratio $\kappa _{eff3D} / \kappa _{eff}$ (dotted red line). The arrows show the assignation of curves with their corresponding $y$-axis.}
		\label{fig6}
	\end{figure}
To gain more insight into the effect of intra-grain scattering we also 
plotted the term (1+$\Lambda /l^{0.75})$ as a function of $l$ with the $y$-axis on the 
left-hand side of Fig.~\ref{fig6}. It can 
be seen that the term varies strongly by almost a factor of 100 for grain 
sizes ranging from 2 nm to 10 $\mu$m, yet this variation cannot be found in 
the effective conductivities shown in Fig.~\ref{fig5}(b). On the $y$-axis of the 
right-hand side of Fig.~\ref{fig6} we plotted the ratio $\kappa _{eff3D} / \kappa _{eff}$. The term 
$\kappa _{eff}$ is given by Eq.~(\ref{eq10}) and $\kappa _{eff3D}$ is that of Eq.~(\ref{eq8}). The ratio is 
between 2 and 3 for a quite extended range of grain sizes (i. e., from 2 nm to almost 1 $\mu$m) with its 
peak value at 100 nm and it reaches a value close to 1 for grain sizes of 10 $\mu$m. 
Thus, for small grains, even though the intra-grain scattering is quite large, the intra-grain phonon mean free path lowers the effective 
conductivity only by a factor of at most $1/3$ due to the role played by Kapitza resistance. 
The role of intra-grain conductivity, which is given by intra-grain scattering, was also analyzed recently in 
\cite{21}, where it was noticed a large discrepancy of 
Kapitza resistance between the value obtained from the fit of experimental data to 
Eq.~(\ref{eq8}) and the results of Kapitza resistance extracted from molecular dynamics simulations \cite{22}. In principles, Kapitza 
resistance would also vary with the grain size. Using planar geometries the intra-grain conductivity as 
well as Kapitza resistance were estimated from molecular dynamics in \cite{21}. As a general trend it was 
found that Kapitza resistance increases with grain size. This trend of increasing thermal boundary 
resistance with grain size is also found by analyzing the experimental data from Refs \cite{9,20}, however 
this increase is offset by the 
geometrical factor $l$.

\section{Conclusions}
\label{D}
In this work we have studied the effective thermal conductivity of both 
nanocrystalline and ultra-nanocrystalline diamond films by EMT approaches. Usually 
the EMT models invoke the approximation of spherical-shaped inclusions. 
These models are extensively used with good results, but they are 
inconsistent with respect to polycrystalline films because the maximal packing 
with spheres is 74{\%}. We formulate an EMT model that can handle in 
principles any shape of the inclusions. Within this model we compare an EMT 
model based on spherical inclusions with an EMT model based on cubic 
inclusions. The results of these two models are very similar. In the 
following we provide a geometrical argument based on the spectral properties 
of the electrostatic operator for sphere and cube \cite{14, 18, 19}. Thus we 
invoke Eq. (\ref{eq7}) and the following sum rules \cite{15,18}: $\sum\limits_k {w_k } = 
1$ and $\sum\limits_k {w_k L_k } = 1 \mathord{\left/ {\vphantom {1 6}} 
\right. \kern-\nulldelimiterspace} 6$, as well as the inequality $0 < L_k < 
1$. For large crystallite sizes, ${R_K \kappa _1 } \mathord{\left/ 
{\vphantom {{R_K \kappa _1 } l}} \right. \kern-\nulldelimiterspace} l \ll 
1$, thus the effective conductivity approaches the bulk value regardless of 
the crystallite shape. On the contrary, for nanometer sized crystallites, 
${R_K \kappa _1 } \mathord{\left/ {\vphantom {{R_K \kappa _1 } l}} \right. 
\kern-\nulldelimiterspace} l \gg 1$, the effective conductivity is $\kappa 
_{eff} \approx \sum\limits_k {{w_k l} \mathord{\left/ {\vphantom {{w_k l} 
{\left( {dR_K L_k } \right)}}} \right. \kern-\nulldelimiterspace} {\left( 
{dR_K L_k } \right)}} $. Knowing that $\left| {1 - 2L_k } \right| < 1$ and 
looking at the spectral parameters {\{}$w_{k}, L_{k}${\}} of both sphere and 
cube, we may perform the following approximations

\[
\sum\limits_k {{w_k } \mathord{\left/ {\vphantom {{w_k } {L_k }}} \right. 
\kern-\nulldelimiterspace} {L_k }} = \sum\limits_k {\frac{2w_k }{1 - \left( 
{1 - 2L_k } \right)}} \approx \sum\limits_k {2w_k \left( {1 + \left( {1 - 
2L_k } \right)} \right)} .
\]
Finally, using the sum rules shown above we conclude that the EMT model 
based on spherical inclusions gives similar results with the model based on 
cubic inclusions. We further notice that both types of inclusions (spherical and cubical) preserve 
macroscopic thermal isotropy of a polycrystalline film.

Moreover, we have found that for grain sizes below 100 nm the effective 
conductivity is strongly affected by both the boundary Kapitza resistance 
and the intra-grain scattering, both of them increasing with grain size. However, the 
increase of Kapitza resistance is offset by the geometrical effect of the grain size.









\end{document}